# Chapter 7

# Machine Protection, Interlocks and Availability

*A. Apollonio, T. Baer, K. Dahlerup-Petersen, R. Denz, I. Romera Ramirez, R. Schmidt, A. Siemko, J. Wenninger, D. Wollmann\* and M. Zerlauth*

CERN, Accelerator & Technology Sector, Geneva, Switzerland

## 7 Machine protection, interlocks and availability

### 7.1 Machine protection with a 700 MJ beam

The combination of high intensity and high energy that characterizes the nominal beam in the LHC leads to a stored energy of 362 MJ in each of the two beams. This energy is more than two orders of magnitude larger than in any previous accelerator. For the HL-LHC it will increase by another factor of two as shown in the comparisons in Figure 7-1. With intensities expected to increase up to $2.3 \times 10^{11}$ p/bunch with 25 ns bunch spacing and $3.7 \times 10^{11}$ p/bunch with 50 ns bunch spacing [1], an uncontrolled beam loss at the LHC could cause even more severe damage to accelerator equipment than at today's nominal beam parameters. Recent simulations that couple energy deposition and hydrodynamic simulation codes show that the nominal LHC beam can already penetrate fully through a 20 m long block of copper if the entire beam is accidentally deflected. Such an accident could happen if the beam extraction kickers deflect the beam at an incorrect angle. Hence, it becomes necessary to revisit many of the damage studies in light of the new beam parameters [2]. In addition, new failure scenarios will have to be considered following the proposed optics changes and the installation of new accelerator components such as crab cavities and hollow electron beam lenses. Special care is required to find a trade-off between equipment protection and machine availability in view of the reduced operational margins (e.g. decreasing quench limits and beam loss thresholds versus increased beam intensity and tighter collimator settings, UFOs at higher energies, reduced bunch spacing, etc.)

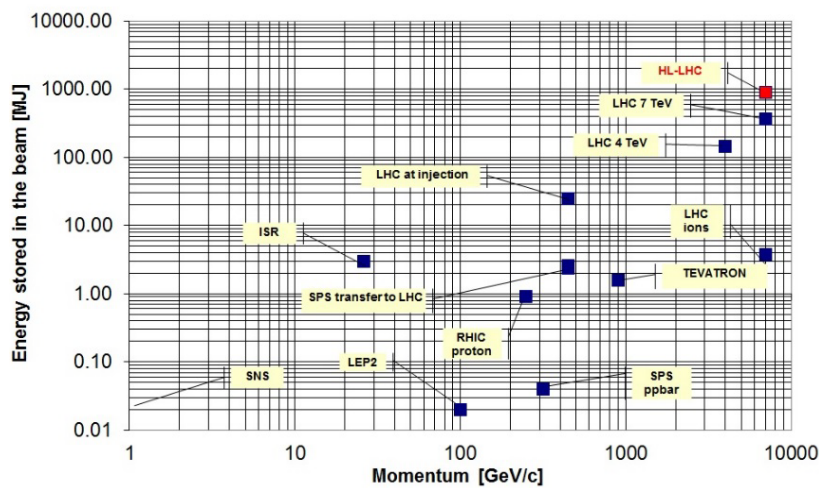

Figure 7-1: Stored beam energy as a function of HL-LHC beam momentum in comparison with other particle accelerators.

---

\* Corresponding author: daniel.wollmann@cern.ch

Safe operation of the LHC currently relies on a complex system of equipment protection. The machine protection system (MPS) is designed for preventing the uncontrolled release of energy stored in the magnet system and damage due to beam losses, with very high reliability. An essential element of the active MPS system is the early detection of failures within the equipment. In addition, the beam parameters are monitored with fast and reliable beam instrumentation. This is required throughout the entire cycle, from injection to collisions. Once a failure is detected by any of the protection systems, the information is transmitted to the beam interlock system (BIS), which triggers the extraction of the particle beams via the LHC beam dumping system (LBDS). It is essential that the beams are always properly extracted from the accelerator via 700 m long transfer lines into large graphite dump blocks. These are the only elements of the LHC that can withstand the impact of the full beams.

The current machine protection architecture is based on the assumption of three types of failure scenarios [3].

- Ultra-fast failures: failures within less than three turns, e.g. during beam transfer from the SPS to the LHC, beam extraction into the LHC beam dump channel, or the effect of missing beam–beam deflection during beam extraction (1 LHC turn = 88.9 μs). In the case of these failures, passive protection elements are required to intercept the beams and protect the accelerator equipment from damage, as no active protection is possible.

- Fast failures: a timescale of several LHC turns (less than a few milliseconds) as a result of equipment failures with a rapid effect on particle trajectories. The active extraction of the beams is completed within up to three turns after the detection of the failure and hence provides protection against such failures.

- Slow failures: multi-turn failures on timescales equal to or more than a few milliseconds, e.g. powering failures, magnet quenches, RF failures, etc.

## 7.2 Protection against uncontrolled beam losses

Equipment failures or beam instabilities appearing on the timescale of multiple turns allow for dedicated protection systems to mitigate their effects on the circulating beams. Figure 7-2 shows a comparison of the failure detection times of different protection systems. As shown in the figure, the LHC beam loss monitoring system (BLM) has the fastest detection time of 40 μs. The BLM system is complemented with fast interlocks on the beam position in IR6, fast magnet current change monitors (FMCM) and a beam lifetime monitor (currently under development by the beam instrumentation group at CERN). All of these systems feature similar failure detection times in the 100 μs to 1 ms range, providing diverse redundancy to the BLM system.

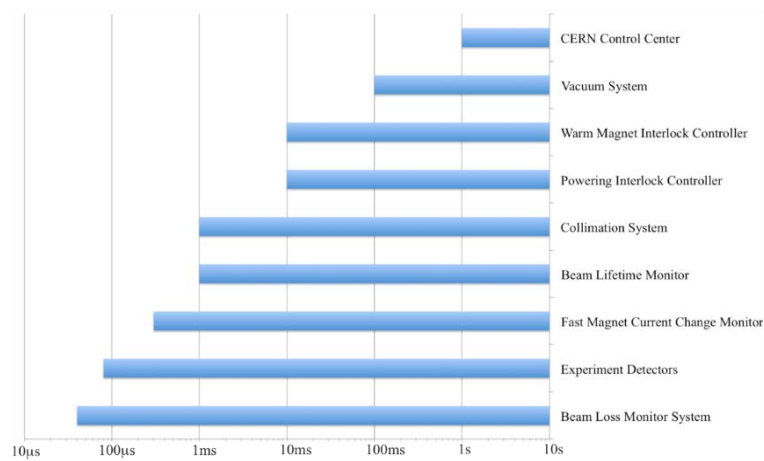

Figure 7-2: Some failure detection times at the LHC. The shortest failure detection time is currently assured by the BLM system, with a fastest integration time of 40 μs, which is equivalent to half a LHC turn.

Adding the additional time required to transmit the detected failure through the LHC beam interlock system, the time required to synchronize the firing of the beam dump kickers with the abort gap as well as the time needed to completely extract the beam from the LHC leads to an equivalent worst case MPS response time of three LHC turns after the failure detection as depicted in Figure 7-3.

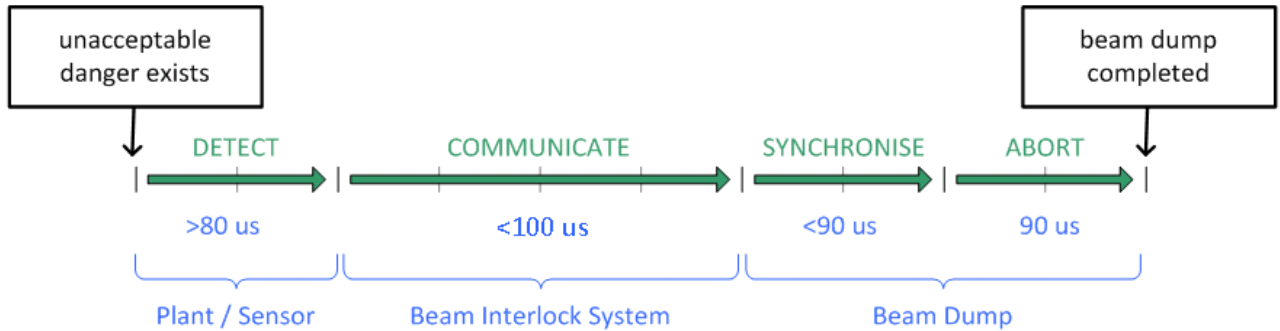

Figure 7-3: Current MPS response time from failure detection to completion of beam dump

This reaction time is sufficient in the absence of failures occurring on timescales below 10 LHC turns. A failure of the normal conducting separation dipole D1 in IP1 and IP5 is currently considered the fastest possible failure with circulating beam. Therefore, this was the basis for the design of the current MPS system. Due to their location in areas with high beta functions and the fast decay of magnet current in the case of a magnet powering failure, these normal conducting magnets can induce fast changes of the particle trajectory. These changes lead to rapidly increasing beam losses in the LHC betatron cleaning insertion (IR7), which define the smallest aperture in the LHC. At nominal energy and intensity the losses after that special failure can reach collimator damage levels within just ten turns. Therefore, a dedicated protection system – the so-called fast magnet current change monitors (FMCM) – has been very successfully deployed on critical magnets in the LHC and its transfer lines in 2006 [4].

With the HL-LHC upgrade, the optics in the insertion regions will significantly change. For certain types of ATS optics the $\beta$-function at the D1 separation dipole magnets in IR1 and IR5 will increase up to ~17 000 m, which will enhance its effect on the beam trajectory. The replacement of the D1 separation dipole magnets by a superconducting magnet would significantly increase the time constants of these circuits, practically mitigating the potential of fast failures originating from these magnets.

For HL-LHC operation, the use of crab cavities will introduce failures that can affect the particle beams on timescales well below the fastest failures considered so far [5]. Studies of different failure scenarios are still underway. These studies require consideration of details of the design eventually to be adopted for the crab cavity and the corresponding low-level RF system. Both have a significant impact on the effect on the circulating beams following, e.g. cavity quenches or trips of the RF power generator. In addition, detailed measurements of the quench and failure behaviour of the chosen design have yet to be conducted. First experience with similar devices at KEK, however, shows that certain failures can happen within just a few turns, as depicted in Figure 7-4.

While protection against failures with time constants >15ms is not expected to be of fundamental concern, voltage and/or phase changes of the crab cavities will happen with a time constant $\tau$, which is proportional to $Q_{ext}$. For a 400 MHz cavity with a $Q_{ext} = 1 \times 10^6$ this will result in a time constant as low as 800 µs. The situation becomes even more critical for cavity quenches, where the energy stored in the cavity can be dissipated in the cavity walls on ultra-fast timescales. Failures believed to be quenches observed in cavities at KEKB show a complete decay of the cavity voltage within 100 µs, accompanied by an oscillation of the phase by 50° in only 50 µs. Such crab cavity failures can imply large global betatron oscillations, which could lead to critical beam losses for amplitudes above about 1 $\sigma_{nom.}$ Highly overpopulated transverse tails compared with Gaussian beams were measured in the LHC. Based on these observations the energy stored in the tails beyond 4 $\sigma$ are expected to correspond to ~30 MJ for HL-LHC parameters. These levels are significantly beyond the specification of the collimation system, capable of absorbing up to 1 MJ for very fast accidental beam losses.

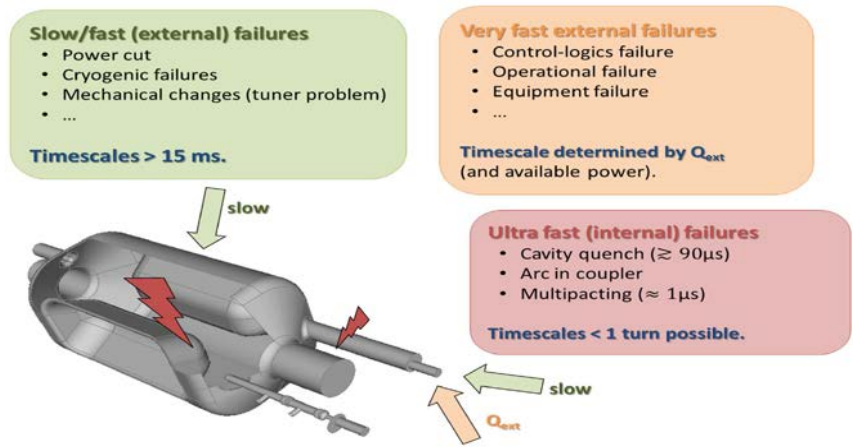

Figure 7-4: Schematic overview of crab cavity failure categories [5]

Therefore, mitigation techniques have to include a fast, dependable, and redundant detection and interlocking of a crab cavity failure on these timescales. Appropriate measures must be taken when designing the cavity and associated RF control to increase as much as possible the failure time constants.

- Avoid correlated failures of multiple cavities (on one side of an IP) through mechanical and cryogenic separation of the individual modules and appropriate design of the low-level RF [6].

- Investigate the use of fast failure detection mechanisms such as RF field monitor probes, diamond beam loss detectors, power transmission through input coupler and head–tail monitors.

- Ensure the partial depletion of the transverse beam tails to reduce the energy stored in the beam halo, which would potentially be deflected onto the collimation system beyond the design value of 1 MJ. For the current baseline this would correspond to an area of 1.7 $\sigma_{nom}$ (before reaching the closest primary collimator) as the possible transverse beam trajectory perturbation following an ultra-fast failure of a single crab cavity. It is important to note that the partial depletion of the beam halo may have a negative effect on the available time to detect a failure with other machine protection systems like BLMs. The consequences of this need to be studied carefully.

- Decrease the reaction time of the MP system for such ultra-fast failures by, e.g. increasing the number of abort gaps, accept the triggering of asynchronous beam dumps with potential local damage, add direct links to the beam dumping system in IR6, and consider the installation of disposable absorbers.

7.2.1   Beam interlock system

The beam interlock system (BIS) is a highly dependable fast interlock system and a key element of the accelerator machine protection. It is currently used in the LHC, SPS, and parts of the injector chain at CERN. Its primary objective is to provide a highly reliable link between users requesting a beam abort and the beam dumping system. The hardware implementation of the system is based on custom-made electronics, as industrial solutions have not been found to be adequate for the specific requirements of the system, e.g. in terms of reaction time combined with the geographical distribution of the system. Due to the obsolescence of electronics components and potential problems with the optical links the present system will need to be upgraded. To fulfil the requirements of the HL-LHC, the system will be equipped with additional input channels to connect more user interfaces and to provide more flexibility in the configuration of the various user inputs. The number of required channels is subject to a future functional specification to be provided by the SPS and LHC machine protection panel (MPP). The possibility of implementing very fast interlock channels and direct links between the crab cavities and the LHC beam dumping system will be studied but the feasibility cannot yet be confirmed. The new system will be equipped with advanced diagnostic features for all optical links allowing pre-emptive maintenance, e.g. in the case of degraded performance due to the enhanced radiation load in some underground areas.

#### 7.2.1.1 Equipment performance objective

The upgraded system is supposed to reach the same performance level in terms of reliability as the present system, which corresponds to safety integrity level (SIL) 3. The safety critical part of the BIS hardware architecture will be based on well-proven principles and solutions but adapted to state-of-the-art electronics components and assemblies. It is therefore probably possible to reuse a major part of the safety critical code, which is very well tested and fully validated. Considering the system's complexity and dimensions, and the experience gained so far, it is expected that system performance will hardly affect LHC availability, e.g. not producing more than one spurious beam abort per year.

The new BIS could be equipped with a new hardware controls interface, replacing obsolete architectures for communication busses and simplifying maintenance and potential upgrades. The BIS hardware will also feature advanced diagnostic tools for the system hardware and the optical links.

All of the proposed changes will require a major revision of the high-level supervision and controls software and be adapted to the accelerator controls environment then in place.

### 7.2.2 Fast magnet current change monitor

The main task of the fast magnet current change monitors (FMCM) is to monitor fast current changes in electrical circuits with normal conducting magnets. A fast current change can be caused by sudden powering failures or perturbations on the supply network, which change the particle trajectories leading to fast beam losses. These monitors are required for electrical circuits with a short decay time constant and magnets installed in regions with high beta functions. Each monitor delivers a permit signal to the beam interlock system to request the extraction of particle beams before losses occur. Therefore, the FMCM provides a redundant protection to the beam loss monitors (BLM). A total of 26 monitors are presently installed to protect the LHC and SPS-LHC transfer lines.

#### 7.2.2.1 Objectives for HL-LHC machine performance

The installation of FMCMs is required to ensure machine protection against powering failures in critical magnetic elements during all operational phases. Twelve monitors are currently installed in the LHC, namely for dump septa magnets in IR6, collimation insertion regions in IR3/IR7, Alice compensator circuits in IR2, and main separation dipoles D1 in IR1 and IR5 [6]. Additional input from WP3 (magnets) and WP2 (accelerator physics) is required to clarify the necessity of additional FMCM units for the protection of the new magnet powering in the high luminosity insertion regions IR1 and IR5 and the HL-LHC optics. In addition, new failure modes derived from the introduction of new elements (such as crab cavities) need to be studied to understand the machine protection requirements and to estimate the number of monitors required to protect the accelerator equipment in the HL-LHC.

#### 7.2.2.2 Equipment performance objective

FMCMs have successfully operated in the LHC and the SPS-LHC transfer lines for many years, and no missed dump has been identified since the start of operation. For the protection of the electrical circuits in the HL-LHC the use of the same design is recommended. However, a review of the system needs in view of the HL-LHC requirements, followed by a corresponding reproduction of additional units, will be required. The review and potential redesign of the hardware is a mandatory step due to the aging and obsolescence of the electronics parts used in the current system.

The aim of the upgrade of the FMCM will be to improve the maintainability of the system and to comply with the requirements of the HL-LHC.

## 7.3 Magnet powering protection

During operation at 7 TeV the energy stored in the main dipole circuits of each sector reaches ~1 GJ. This illustrates that the LHC magnet system needs to be protected against damage due to failures in magnet powering or quenches.

Therefore, the superconducting circuits, busbars, and future superconducting links must be equipped with a quench protection system (QPS) that detects changes in the superconducting state and activates quench heaters and/or the energy extraction systems (EE) to safely extract the magnetic energy stored in the circuit.

Furthermore, the correct powering conditions have to be ensured for each circuit by a powering interlock system (PIC), which will interlock the powering of the circuit via the power converter in the case of problems and potentially request a beam dump. Therefore, the PIC is interfacing the quench protection systems, power converters, cryogenic systems, and technical services such as uninterruptable power supplies (UPS), emergency stop buttons (AUG), and controls.

The protection of the normal conducting magnets in the LHC and its injector complex is ensured by the warm magnet interlock system (WIC), which collects signals from thermo switches installed on the magnets and status signals from the associated power converters.

As a failure in the magnet system will also impact the stored particle beam, these systems have to interface the magnet powering systems with the BIS and initiate a beam dump in case of a failure.

### 7.3.1 Quench protection system

The HL-LHC will incorporate new superconducting elements requiring dedicated protection systems. The upgrade of the QPS will provide this functionality, including the related data acquisition systems, monitoring the state of the protection systems and the protected elements.

The enhanced luminosity of the HL-LHC will increase the radiation levels in certain underground areas like the dispersion suppressors to levels no longer compatible with the operation of radiation-tolerant electronics based on COTS currently installed in those areas. Based on the progress in electronics, it is probably feasible to re-locate a major part of the protection electronics to low radiation zones or, eventually, to surface buildings and use long instrumentation cables or optical fibres to link to the protected elements.

The proposed upgrade will also include new communications links for supervision and data acquisition (DAQ) superseding the then obsolete classic fieldbus networks. At the same time, advanced tools for remote diagnostics and maintenance will be provided.

#### 7.3.1.1 Equipment performance objective

The QPS is a highly complex system incorporating a large amount of electronics components and assemblies. As for the existing system, a particular design effort will be necessary to achieve the very demanding level of system dependability required for successful LHC operation. The number of accesses to LHC underground areas needs to be minimized for personnel protection and machine availability.

The protection parameter settings will be subject to a functional specification to be issued by the LHC magnet circuits, powering and performance panel (MP3) based on the input of equipment specialists.

The request for very fast magnet protection systems with reaction times in the order of some milliseconds will be addressed in a feasibility study, but a potential implementation will rely strongly on the proper instrumentation of the protected element (magnet, superconducting link) and the adapted reaction time of other protection systems such as energy extraction systems and the powering interlock system.

The hardware capabilities of the DAQ and related communication links will be enhanced to allow higher data transmission rates and advanced maintenance. All of the proposed changes will require a major revision of the high-level supervision and controls software, which also needs to be adapted to the accelerator controls environment then in place.

7.3.2    Energy extraction system

Energy extraction systems (EE) are an integrated part of the safety-critical quench protection systems, which are widely used in the existing LHC machine with a total of 234 installed facilities (202 for the 600 A corrector circuits and 32 13 kA systems for the main dipole and quadruple circuits). The systems are strongly circuit-specific, tailor-made for a particular set of requirements, and adapted to the local infrastructure.

The current mechanical energy extraction switches will have to be replaced for the HL-LHC era. To allow for faster reaction times, which may be required by future superconducting magnets, the development of fast switches based on semiconductors (IGCTs, IGBTs) is currently ongoing. Alternatively, slower (~20 ms) hybrid systems, where an electromechanical switch is connected in parallel to a semiconductor switch, are under study. Furthermore, for slow switching times, classical mechanical switches in combination with large snubber capacitor banks could be used.

The new design of the extraction resistors (DQR) will be significantly different from those developed and built for the present LHC main circuits. The pursued characteristics and properties of these new energy absorbers are very fast recovery (cooling) times, compact design (minimized volume), and easily changeable resistance values.

#### 7.3.2.1    Equipment performance objectives

The new energy extraction equipment for the HL-LHC will use a new generation of switches, incorporate the newest technology for high-current transmission, benefit from built-in features for facilitating diagnostics and maintenance, offer systems that will minimize intervention time for accessing all parts of the facilities, and profit from the experience gained with the operation of the existing LHC EE facilities.

7.3.3    Powering interlock system

The powering interlock system (PIC) ensures the correct powering conditions for the different electrical circuits with superconducting magnets in the LHC. At the same time, it guarantees the protection of the magnet equipment by interfacing quench protection systems, power converters, cryogenics, and technical services such as uninterruptable power supplies (UPS), emergency stop buttons (AUG), and controls. The PIC is a distributed system currently consisting of 36 individual powering interlock controllers, which manage the powering of each of the 28 powering subsectors [7].

#### 7.3.3.1    Objective for HL-LHC machine performance

Magnet interlocks are required to guarantee safe magnet powering during all phases of operation from injection to collisions. In order to achieve this protection while maintaining the time constraints required for equipment protection, interlock electronics are usually installed close to the main clients (QPS and power converters) such as the UA, UJ, and RR alcoves. At the design luminosity for HL-LHC ($5 \times 10^{34}$ cm$^{-2}$ s$^{-1}$) the thermal neutron and high-energy hadron fluencies in the areas close to the tunnel will increase considerably with respect to the values for which the existing PIC has been designed. Additional changes and requirements from the new quench protection system will have to be reflected in the upgraded interlock system to assure the dependability of the system during the HL-LHC era. For these reasons, a new design for the distributed I/O modules is required to cope with the increment of particle flux in the most sensitive areas. In addition, an upgrade of the industrial components used will very likely become necessary due to the changing of the low-level I/O components used.

#### 7.3.3.2    Equipment performance objectives

The PIC was installed during 2006 and has been operating since the start of LHC operation in 2008. By the time the HL-LHC starts operation the system will have been running for more than 15 years. Therefore, some of the electronics components will reach the end of their life expectancy, which can have an impact on the availability of the system. In addition, obsolescence of electronics parts needs to be addressed since some of the critical components of the system can no longer be purchased on the open market, affecting its maintainability. Furthermore, the increased radiation dose in certain areas will affect the most sensitive components of the PIC.

The upgrade of the PIC will address the issues mentioned above to adapt to HL-LHC requirements. In addition, the system will be reviewed to provide enhanced diagnostics of the safety-critical hardwired loops in line with the upgrades foreseen for the quench protection system. The implementation of triple redundancy in combination with two-out-of-three voting is considered for the quench loop.

### 7.3.4 Warm magnet interlock system (WIC)

The warm magnet interlock system (WIC) assures the protection of normal conducting magnets in the LHC and its injector complex. It collects signals from thermo switches installed on the magnets and status signals from the associated power converters. Based on these input signals a programmable logic controller (PLC) calculates and transmits fast abort signals to interlock the power converter of a given electrical circuit in case of powering failures. In addition, it initiates a beam dump in case a circuit relevant for beam operation of the given machine is not operating in nominal conditions [8].

#### 7.3.4.1 Objective for the HL-LHC machine performance

Magnet interlocks are required to guarantee safe magnet powering during all phases of operation from injection to collisions and to abort beam operation avoiding inevitable beam losses in case of powering failures. In addition, interlock systems provide remote diagnostic features to allow an efficient and precise identification of faulty equipment. As normal conducting magnets in the LHC are concentrated around the IPs, a single industrial controller installed in a radiation-free area is required to manage the protection of powering equipment in a given IR. The main objective for the HL-LHC era is a consolidation of the existing system, along with an upgrade to accommodate changes and new requirements for the magnet powering system in the different insertion regions.

#### 7.3.4.2 Equipment performance objectives

The WIC – consisting of eight industrial controllers and several I/O crates – has been installed in the LHC since 2006 and has been continuously operating since the start of LHC operation in 2008. No failures have been observed in the WIC system throughout the whole of Run 1. Nevertheless, by the time the HL-LHC starts operation the WIC system will have been running for more than 15 years, requiring an upgrade of the electronics and industrial components to assure the current level of dependability throughout the full HL-LHC period. Furthermore, the increased radiation dose in certain areas will affect the most sensitive components of the WIC (mostly the magnet and remote test interconnection boxes) for which a new design and production has to be envisaged.

### 7.4 Availability requirements to achieve HL-LHC goals for integrated luminosity

The challenging goals in terms of integrated luminosity require a high level of accelerator availability and operational efficiency. The estimated integrated luminosity as a function of machine availability is analyzed below. There are three important figures to be considered when evaluating LHC availability.

- The stable beams time: the time for beam collisions. This quantity must be optimized by operators as a function of the observed distributions of turnaround time and fault time. Stable beam optimization will be particularly relevant for levelled operation.
- The turnaround time: the time to go from a beam dump at the end of stable beams to the next stable beams, when no faults occur. This quantity has a minimum value imposed by the injection process and the ramp time of superconducting magnets. Efforts must be devoted to reaching this lower value. Parallelizing/combining machine modes (e.g. collide and squeeze) is one possibility to optimize the turnaround time.

- The fault time: the time to clear machine faults and recover operational conditions for beam injection. It includes the time for expert diagnostics and intervention, eventually requiring access to the LHC tunnel.

For the following extrapolations, the 2012 LHC run is taken as a reference, being the most stable and reproducible year of the first LHC run [9]. In 2012 the average turnaround time was 5.5 h, but the energy was limited to 4 TeV, requiring current ramps of only 13 min. For HL-LHC operation, the reference energy will be 7 TeV, resulting in an increased average turnaround time, estimated to be 6.2 h. Making predictions of the fault time distribution for HL-LHC operation is not trivial, as many factors will play a role in this respect. The increased energies and intensities will result in higher radiation levels, which could have a direct impact on the observed number of single event effects (SEE) per year and on long-term effects on components due to the total integrated dose (TID). Mitigations to these effects have been deployed during LS1, with a major relocation of sensitive equipment in protected areas of the LHC tunnel. Nevertheless, future radiation levels will have to be measured for a final assessment of the expected increase in beam dumps due to radiation effects in electronics components (R2E). It is already foreseen that a new generation of electronics systems will be designed before the HL-LHC era. New designs should cope with such radiation levels. The increased energy and intensities will also have a direct impact on the observed UFO-induced beam dumps due to localized losses. Current extrapolations assume a significant increase of UFO events that are large enough to provoke a beam dump, if 2012-like BLM thresholds are to be kept. A balance between the tolerated number of UFO dumps and the possibility of having beam-induced quenches should be found, by the definition of suitable threshold and, eventually, BLM relocation. On the equipment side, components will operate closer to design limits and partially reach their end-of-life.

A Monte Carlo model [10, 11] of LHC availability was used to qualitatively assess the combined impact of all these factors. A sensitivity analysis has been carried out with respect to the average fault time and machine failure rate, defined as the ratio between fills to stable beams followed by failures and the total number of physics fills. The results for 200 days of HL-LHC operation are presented in Figure 7-5. Nominal HL-LHC parameters ($5 \times 10^{34}$ $s^{-1}$ $cm^{-2}$ levelled luminosity, $2.19 \times 10^{35}$ $s^{-1}$ $cm^{-2}$ virtual peak luminosity, 4.5 h average luminosity lifetime, 7 TeV) and the same duration of intensity ramp-up as 2012 were assumed.

The results show that, assuming 2012 figures for average fault duration (6.9 h) and machine failure rate (70%), 260 $fb^{-1}$ could be produced in 200 days of HL-LHC operation. To reach the goal of 300 $fb^{-1}$ per year, a reduction of 10% of the machine failure rate, combined with a reduction of the average fault time of around 25%, are necessary (HL1). As an alternative for reaching 300 $fb^{-1}$, the sensitivity analysis shows that a further reduction of the machine failure rate to 50% would keep the same average system fault time as observed in 2012 (HL2). These requirements in particular need to be considered during the design of future electronics equipment.

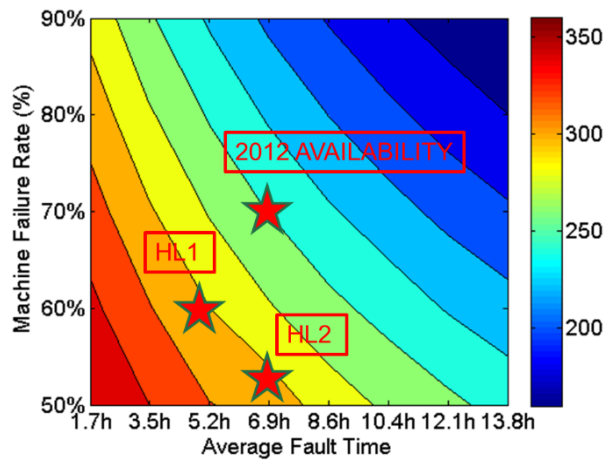

Figure 7-5: Sensitivity analysis of integrated luminosity to the machine failure rate and average fault time.

## 7.5 References

[1] O.S. Bruning, HL-LHC Parameter space and scenarios, Proc. Chamonix 2012 workshop on LHC Performance, CERN-2012-006 (2012) p. 315.

[2] N.A. Tahir, J.B. Sancho, A.Shutov, R.Schmidt and A.R. Piriz, Impact of high energy high intensity proton beams on targets: Case studies for Super Proton Synchrotron and Large Hadron Collider, *Phys. Rev. ST accel. Beams* **15** (2012) 051003.

[3] R. Filippini, B. Goddard, M. Gyr, V. Kain, R. Schmidt, J. Uythoven and J. Wenninger, Possible causes and consequences of serious failures of the LHC machine protection systems, 9th European Particle Accelerator Conf., Lucerne, Switzerland, 5–9 July 2004, p. 620.

[4] M. Werner, Einrichtung zur Bestimmung der Starke des Magnetfeldes eines Elektromagneten, Patent DE102005045537B3 28.12.2006, September 2005.

[5] T. Baer et al., Very fast crab cavity failures and their mitigation, Proc. IPAC'12, May 2012, pp. 121–123, MOPPC003.

[6] M. Werner, M. Zerlauth, A. Dinius and B. Puccio: Requirements for the fast magnet current change monitors (FMCM) in the LHC and SPS-LHC transfer lines, LHC-CIW-ES-0002, EDMS 678140.

[7] R.Schmidt, B.Puccio, M.Zerlauth: The hardware interfaces between the powering interlock system, power converters and quench protection system', LHC-D-ES-0003, EDMS 368927.

[8] P. Dahlen, R. Harrison, R. Schmidt and M. Zerlauth, The warm magnet interlock system for the LHC ring, LHC-CIW-ES-0003, EDMS 653548.

[9] B. Todd *et al*., Workshop on Machine Availability and Dependability for Post-LS1 LHC, Proc. Workshop, CERN (2013).

[10] A. Apollonio *et al*., HL-LHC: Integrated luminosity and availability, TUPFI012, Proc. IPAC'13 Shanghai, China, 2013, 1352-1354.

[11] A. Apollonio *et al*., Update on predictions for yearly integrated luminosity for HL-LHC based on expected machine availability, TUPRO015, Proc. IPAC'14, Dresden, 1036-1038.